Review

# Extended Reality for Mental Health Evaluation —A Scoping Review

Olatunji Omisore, PhD[1]; Ifeanyi Odenigbo, MSc[2]; Joseph Orji, MSc[2]; Amelia Beltran, MSc[2]; Meier Sandra, PhD, MD[3]; Nilufar Baghaei, PhD[4]; Rita Orji, PhD[2]

[1]Shenzhen Institute of Advanced Technology, Chinese Academy of Sciences, Shenzhen 518055, China.

[2]Faculty of Computer Science, Dalhousie University, Halifax, Canada

[3]School of Information Technology and Electrical Engineering, University of Queensland, Brisbane, Australia

[4]Department of Psychiatry, Dalhousie University, Halifax, Nova Scotia, Canada

**Corresponding Author:**
Olatunji Omisore,
Shenzhen Institutes of Advanced Technology,
Chinese Academy of Sciences, Shenzhen 518055, China.
Phone: 8613172482240
Email: ootsorewilly@gmail.com

## Abstract

**Background:** Mental health disorders are the leading cause of health-related problems globally. It is projected that mental health disorders will be the leading cause of morbidity among adults as the incidence rates of anxiety and depression grows globally. Recently, extended reality (XR), a general term covering virtual reality (VR), augmented reality (AR) and mixed reality (MR), is paving a new way to deliver mental health care. In this paper, we conduct a scoping review of the development and application of XR in the area of mental disorders.

**Objective:** We investigated the adoption and implementation XR technologies used for the intervention of mental disorders. This study aims to provide statistical analyses of the design, usage, and effectiveness of XR technology for mental health intervention with a global demographic focus.

**Methods:** We performed a scoping database search to identify the relevant studies indexed in Google Scholar, PubMed, and the ACM Digital Library. A search period between August 2016 and December 2023 was defined to select articles related to the usage of VR, AR, and MR in a mental health context. The database search was performed with pre-defined queries, and a total of 831 articles were identified. Ten (10) additional articles were identified through professional recommendation. Inclusion and exclusion criteria were designed and applied to ensure that only the relevant studies were included in the literature review.

**Results:** We identified a total of 85 studies from 27 countries across the globe. By performing data analysis, we found that most of the studies focused on developed countries such as the United States (16.47%) and Germany (12.94%). None of the studies were for African countries. The majority of the articles reported that XR techniques led to a significant reduction in symptoms of anxiety or depression. The majority of studies were published in the year 2021, i.e., 31.76% (n = 31) of the included studies. This could indicate that mental disorder intervention received a higher attention when COVID-19 emerged. Most studies (n = 65, 76.47%) focused on a population between the age range of 18 to 65 years old, while only fewer studies focused on teenagers (n = 2, 3.35%), i.e. subjects between the ages of 10 to 19 years old. Also, more studies were done experimentally (n = 67, 78.82%) rather than by analytical and modeling approaches (n = 8, 9.41%). This shows that there is a rapid development of XR technology for mental health care. Furthermore, these studies showed that XR technology can effectively be used for evaluating mental disorders in similar or better way as the conventional approaches.

**Conclusions:** In this scoping review, we studied the adoption and implementation of XR technology for mental disorder care. This covers 85 studies that used different types of VR, AR, and MR techniques for managing 14 types of mental disorders. Our study identifies that XR treatment yields high patient satisfaction and follow-up assessments show significant improvement with large effect sizes. Moreover, the studies adopted unique designs that are set up to record and analyze the symptoms reported by their participants. This review study may aid future research and development of various XR mechanisms for differentiated mental disorder procedures.

# Introduction

## Background

Mental disorders are defined as behavioral or mental patterns that cause significant distress or impairment for an individual. These are highly prevalent and, currently, are the leading cause of disability globally. In the last decades, a global increase in the incidence of these disorders has been observed [1, 2]. According to the World Health Organization (WHO), mental disorders are the leading cause of disability in the United States and the United Kingdom. The WHO predicted that mental disorders would account for 13% of the total burden of diseases by 2030 [3]. As an indication, around 20% of adults experience one type of mental health problem in the United States, United Kingdom, and related developed countries [4]. A recent survey shows that acceleration of socioeconomic developments increases the prevalence of mental disorders (17.5% adults) in China [5]. Meanwhile, adolescents are characterized with the highest incidence of mental disorders in Canada [6]. While over 75% of people with mental disorders remain untreated in developing countries, 35 - 50% of the corresponding range is also found in high-income countries [7].

The WHO estimated that these mental conditions will cost the global economy about $1 trillion in lost productivity annually [8]. Some of the mental healths are interlinked. For instance, anxiety and depression remains the most common mental disorders in today's society [9]. Anxiety is closely linked to mood disorders, and individuals developing depression would often experience anxiety disorder at some earlier point(s) of life [10]. While the etiology of anxiety disorders and depression is complex, multiple causal factors such as rapid social change, stressful work conditions, gender discrimination, social exclusion, unhealthy lifestyle, physical ill-health, human rights violations, as well as genetics have been appropriately studied. Many times, mental health researchers have studied the positive effects of evaluating anxiety in combination with and other mental conditions like pain or depression. For example, Bandelow *et al.* [11] reported that 1 out of every 13 mental disorders is anxiety with major depressive and specific phobia disorder. In general, reports mostly suggested that closer mental care should be addressed by increasing the accessibility and the development of tools that patients can use on their own [3].

## Conventional Assessment Approaches

Cognitive Behavioral Therapy (CBT) is a conventional approach that was shown to be effective in the treatment of a wide range of mental disorders such as anxiety disorders, depression, phobia, and alcohol use problems [12]. CBT is based on the core principles that thoughts impact feelings and feelings impact behavior. During CBT patients learn to change maladaptive thinking patterns and novel coping behaviours to become and stay healthy. CBT can be as effective as, or even more effective than, other forms of psychological therapy or even psychiatric medications especially for patients diagnosed with anxiety disorders or depression. CBT is well supported by many clinical practice guidelines [12]. Studies have shown that it is an evidence-based therapy that reliably helps in overcoming depression. However, it involves aiding people to identify and change the bad lifestyles that negatively influences their behavior and emotions [13]. Rather than being a set method, CBT combines procedures that are developed on a certain disorder that has been unevaluated. For instance, the treatment procedure for depression is different from how CBT is used in evaluating phobia and other anxiety disorders.

Exposure therapy is a major element of CBT that is more focused on certain mental healths related to anxiety [14]. In this approach, participants, the subject been assessed, are exposed to feared objects, activities or situations in a safe environment, and this is known to reduce patients' fear and possibility of avoidance. With gradual follow-up, participants learn to overcome their anxiety [15]. The variations of exposure therapy can be majorly classified as conventional and modern exposure, usually based on the application context. Conventional exposure includes both *in-vivo* and imaginal exposure. During *in-vivo* exposure, patients are intentionally faced with real-world objects or situations they feared to reduce their anxiety [16]; however, it only works for a small percentage of mental health cases. In contrast, imaginal exposure configures an alternative approach during which patients imagine the worst outcome scenarios to confront their fears within their mind. The effectiveness of imaginal exposure depends on a patient's motivation, and their ability to generate fear-inducing imaginations. Exposure therapy is challenging as therapists require extensive training and multiple long exposure sessions. Consequently, the conventional methods are time-consuming and costly. Recently, XR are evaluated as a new approach for delivering exposure-based therapy for mental disorders. The potentials of XR for treating anxiety and depression have been reported [17, 18].

## Use of Extended Reality in Mental health

Extended Reality (XR) is an umbrella term referring to all real, virtual, and mixed environments, wherein interactions are generated by computer technology to engage humans [14]. XR is a rapidly growing technology and is playing prominent roles in different sectors such as providing clear benefits in many aspects of work and business, including training, collaborative working, and marketing. The technology is rapidly gaining traction in creating imagination of real worlds through virtual, augmented, and mixed realities. XR was recently conceived for carrying out evaluation of mental healths. Thus, patients suffering from mental health can be virtually immersed to allow them to display and confront the disorders they suffer. It has been noted that the advances in XR tools can transform the health domain remarkably; however, an exciting issue is studying the adoption and implementation levels of current VR, AR, and MR techniques for evaluating mental health [19-21]. In the industry sector, recent reports shows that the XR medical market is estimated to reach 1.7 billion USD in 2022, with a compound annual growth rate of 105.6% from 2018 to 2022 [22]. Thus, supporting XR-based solutions will play a crucial role in the future of mental health. As the market continues to grow, it is safe to assume that developing XR technologies for mental disorder interventions will continue to increase.

In mental health interventions, XR techniques involve the use of single or multiple base technologies to create exposures. The base technologies namely virtual reality (VR), augmented reality (AR) and mixed reality (MR) involves using computer models to artificially design real world environments with stimuli sensory features. Thus, the environment can simulate typical contexts that induce mental healths such as anxiety, phobia, or pain to enable users to interact with the environment. Typically, the artificial environment can be developed using four main components which are:

1. a high-end graphics rendering unit that is used to compute and render virtual scenes via frame buffer;
2. a 3D stereo display unit that connects users' visual sensory system to the environment;
3. a tracking system that models users' movement in the virtual environment; and,
4. other input interfaces such as joysticks or sensory gloves which provide tactile feedback.

Currently, existing studies suggest that XR-based evaluations can be as effective as conventional exposure-based methods [12, 23-25]. It is anticipated that XR technology will offer the greatest promise for mental health care [12]. This is because XR-based exposure therapies are found to be very accessible and can offer lasting improvements for different mental health conditions. By analyzing many existing studies, we found that a good number of XR techniques exist. These are used to evaluate different mental disorders via different software and hardware technologies [12, 23, 26, 27]. XR systems have been successfully applied in individual, group-based, and internet-based intervention of mental healths [28-30]. The adoption of XR systems started around two decades ago when Hoffman's team [31] developed a VR gaming system for exposure-based therapy in mental care, called *SnowWorld*. The *game* provides a systematic way of reducing players' pain perception during burn wound care. Anderson *et al*. [23] presented a follow-up of the first randomized clinical trial to test another format for delivering CBT for social anxiety disorder—virtual reality exposure therapy. The study showed that VR and exposure group therapy has been well established as an effective strategy for evaluating social anxiety disorder.

The application of XR technologies for mental health care delivery provides opportunities for customizing, reproducing, and tweaking several evaluation parameters according to an individual patient's needs during mental health care. VR or AR environments provide a greater degree of control for therapists to customize, reproduce, and tweak several evaluation parameters according to an individual patient's needs. Such parameters include fan wind, stereo sound, moving chair, color display, and odor emitters [32]. This kind of customization may not be achieved in traditional exposure therapy [33, 34]. Also, the risks associated with privacy intrusion are reduced as everything is transformed into a virtual environment [35]. Simulated and augmented environments are less scary than the use of in-vivo and imaginal exposure in conventional therapy [31]. Exposure-based therapies defined on VR/AR/MR applications have been shown to be effective for evaluating different mental health conditions. This study presents findings of a scoping review on the state-of-the-art XR systems used in mental disorder care.

### Objective

Recently, XR-based mental health intervention is advancing rapidly. It is critical to analyze the implementation and adoption levels of the state-of-the-art XR techniques used for mental care delivery. This scoping review was carried out to investigate the implementation and adoption levels of XR (i.e., a combination of studies that reported VR, AR, and MR) for mental health care delivery. The review was conducted following the guidelines outlined by Arksey and O'Malley [36]. The main objective of this scoping review is to show the implementation and usage levels of XR-based therapy in providing care for different mental disorders across the globe. Thus, this review study is set to provide a statistical analysis of studies that have recently focused on: ***a)*** technological design and usage of XR in mental health with a global demographic focus; ***b)*** components that are found in different XR interventions used for mental disorders; and ***c)*** effectiveness of the XR technology for anxiety and depression as top mental disorders.

## Methods

### Eligibility Criteria

The adoption of VR, AR, and MR for mental disorder evaluation has evolved over time. The rapid advancement happens in a corresponding timeline with developments in the hardware and software used for implementing the XR technologies. Hence, we decided to limit our data sources to articles published between August 2016 and December 2023 so as to analyze the state-of-the-art in the study area. Advance search sections of the three databases by the authors OM, IO, JO, AB individually and the articles located were later combined. This is to ensure a wide coverage of articles published in the search period. Further, only a limited set of search criteria were used to limit the articles extracted to more relevant ones. We only considered the studies that were published in peer-reviewed journals and refereed conferences (with oral presentation). Ten (10) supplementary records were also identified via professional sourcing. Based on our search strategy and study goal, we decided to use a combination of two search rules namely: ***i***) all the search terms must be present in the article' title and/or abstract, and ***ii)*** the article publication year must be within the specified range between August 2016 and December 2023. Additionally, exclusion criteria were defined as all ***i)*** duplicate articles, ***ii)*** version updates, ***iii)*** articles written in language other than English, ***iv)*** studies that reported anxiety or depression as a secondary aspect or induced illness, and ***v)*** articles presenting SWOT analysis, thesis and citations, and scoping reviews, were filtered out.

### Information Sources

We formulated a search strategy used to explore multiple databases to find all recent and relevant studies about XR technologies. We focus on information from studies that focused on depression and anxiety and related mental disorders. The scoping search was done on three different databases which are library sources for research articles, grey literature, patents and common information. Our choices of databases are: (1) PubMed[a] (2) Google Scholar[b] and (3) the ACM Digital Library[c]. The databases were chosen because they provide an interface to generate wild search queries across a variety of disciplines, databases, and journals. In addition, they have the most complete indexes of articles that focus on the theme of this review study. These aided to simultaneously access a broad range of evidences including technical and peer-reviewed studies reported from different all parts of the world, different publishers, and over a long time period. We defined our search period to filter out only articles published between August 2016 and

[a] https://pubmed.ncbi.nlm.nih.gov/advanced/),
[b] https://scholar.google.com
[c] https://dl.acm.org/search/advanced

December 2023 and indexed in any of the three databases. Overlapping articles were filtered to avoid duplication. Multi-level filtering was carried out following the Preferred Reporting Items for Systematic Reviews and Meta-Analyses (PRISMA) guidelines [36] and associated checklist in the *Supplementary Materials*. This was done to limit the search outcomes to relevant studies that can provide the most valuable data to answer our research objective. The search strategy was set to limit data sources to studies that implemented or used VR, AR or MR for different mental health care.

## Search Strategy

Our article search strategy is based on using an organized structure of search terms to retrieve existing literature in the database. We combined the keywords in the objectives of this review study in order to retrieve relevant articles from the databases. The search terms were discussed amongst the research team and were defined as "Augmented Reality", "Mixed Reality", "Virtual Reality", "Depression", and "Anxiety", "Mental Health". We choose the free-text search as it is more flexible that we could target the free words in both the title and abstract fields to limit the final sets provided an efficient way to increase the specificity our search. These keyword searches were carried out for each concept in our research objective, and we designed the search queries to include a combination of Boolean operators "OR" and "AND" to reduce omission of vital articles. These search terms are the most appropriate keywords that reflected in the subject area and have utmost relevance to our review objective. The full terms were exclusively used during the search to avoid any potential conflicts with other terms such as VR intended for "Virtual Reality" extracting use of related acronyms (e.g., Voice Recognition) which would make the filtering cumbersome and not necessarily generate additional useful resources. The selection criteria were carefully designed to consider articles that contain one or multiple search terms in the study's title and/or abstract sections. The example shows the study's title and/or abstract sections.

## Study Selection

The search strategy yielded 831 articles scrapped from the three databases, and 10 additional articles were identified through a professional source. Specifically, by using the defined search filter criteria "anywhere in the article", the search results included 608, 172, and 51 articles scrapped from PubMed, Google Scholar, and ACM Digital Library databases, respectively. The articles were scrapped and processed by following the set of items in the PRISMA checklist. The checklist and a list of the articles used for this study can be found in the *Supplementary Materials*. First, authors IO, JO, AB independently screened the articles retrieved while author OM performed quality check on all the final records. Next, irrelevant and duplicate articles (*i.e.,* 324 articles) were removed; thus, a total number of 517 articles were left. The remaining articles were further screened for relevance. This started with title screening, irrelevant articles were removed. A total of 176 texts were removed in this step. Full abstract reviews were done in situations when an article's relevance could not be resolved from the title. Thus, 103 articles were further screened out. Yet, authors carried out a review of the full text when certainty on an article's relevancies was still lacking to decide whether they are relevant or not. In total, 279 non-relevant articles were screened out by the authors leaving only 238 articles sought for retrieval. A second screening step was required to limit the scoping review to articles that fulfill the eligibility criteria. Thus, further assessment was carried out and another 127 articles were removed. Full texts of 111 articles were retrieved. Articles that were out of context (n = 20) and those that lacked quantitative data (n = 6) were also excluded. Finally, a total of 85 articles that meet the eligibility criteria were included in this review study.  These procedures were done with Microsoft Excel and without any form of automation in the process.

## Data collection process and information extraction

Three of the authors performed the data extraction process while the data validity and accuracy were checked by a fourth author. The full texts of the 85 articles were downloaded and shared amongst the authors for review. Specific details of the articles listed below were extracted and processed in a spreadsheet to analyze the adoption of VR, AR, and MR in managing anxiety and depression and related mental healths.

1. Publication authors, year, and regions;
2. The study type and study design focus and health domain;
3. Methodology (e.g., study duration, number of sessions, and duration in minutes);
4. The methodology that the study was based on and the evaluation strategy;
5. The VR/AR/MR application and technology (type of headset, toolkit) used for the study;
6. Study demographics such as targeted population, sample size, and age distribution;
7. The motivational strategies, targeted outcome, and region;
8. Key findings on using the XR techniques for managing depression and anxiety.

The details related to the abovementioned data were utilized to address the specific research objective that is guiding this scoping study. The data provided useful insights could help developers and researchers on VR/AR/MR with the future research. Also, users can learn the important of such systems like the use of XR-based exposure therapy for their mental health.

## Results

### Publications Demographics by Country and Year

This scoping review was based on a total of 85 articles [28, 29, 35, 37-121] which are attached as part of *Appendix 1.* We reported statistics and meta-analysis of studies that addressed the technological design and usage of XR in mental health, major components used in the different XR interventions for management of the mental disorders; and effectiveness of the XR technology for anxiety and depression as top mental disorders. The article selection procedures strictly follow the steps discussed in Figure 1.

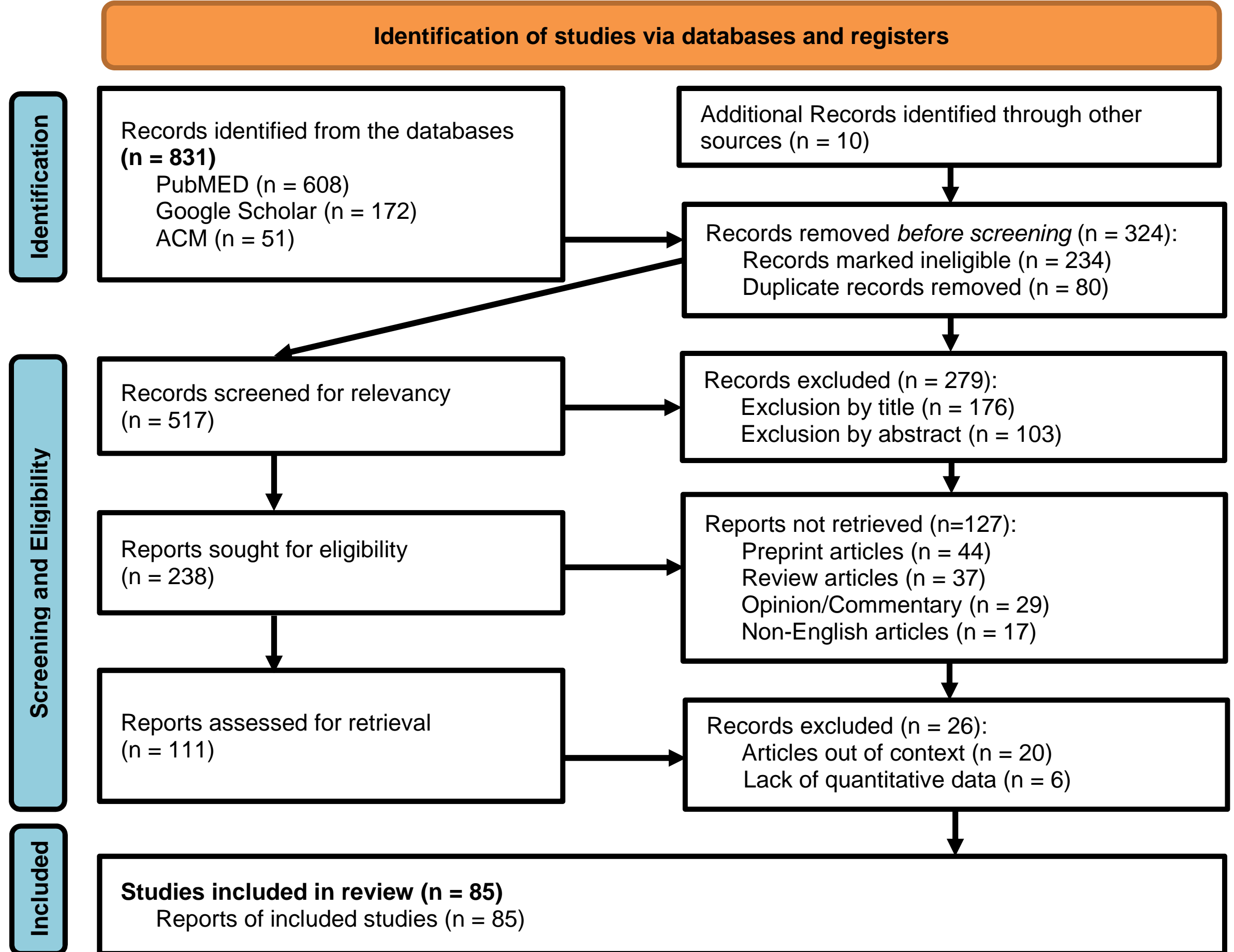

**Figure 1.** Literature screening and selection flowchart following PRISMA guidelines [36]

***Table 1***. *List of countries that conducted mental health studies with the XR technologies*

| SN | Country of Study | Number of Studies | SN | Country of Study | Number of Studies |
|---|---|---|---|---|---|
| 1 | Armenia | 1 | 15 | Jordan | 2 |
| 2 | Australia | 2 | 16 | Korea | 8 |
| 3 | Austria | 2 | 17 | Netherland | 6 |
| 4 | Belgium | 1 | 18 | Philippines | 1 |
| 5 | Canada | 1 | 19 | Poland | 2 |
| 6 | China | 5 | 20 | Portugal | 2 |
| 7 | Denmark | 2 | 21 | Romania | 2 |
| 8 | France | 1 | 22 | Singapore | 1 |
| 9 | Germany | 11 | 23 | Spain | 7 |
| 10 | Hong Kong | 1 | 24 | Sweden | 1 |
| 11 | India | 1 | 25 | Turkey | 1 |
| 12 | Iran | 3 | 26 | United Kingdom | 4 |
| 13 | Israel | 2 | 27 | United States | 14 |
| 14 | Japan | 1 | | | |

First, we analyzed the country of origin of the articles. The 85 studies have been conducted in 27 countries across the globe, as presented in Table 1. 16.47% (n = 14) of the studies were carried out in the United States, and this is followed by Germany (n = 11, 12.94%). Compared with the previous study [13], it may be said that both countries dedicated a good amount of research funding and time to study how XR aid mental health care in the United States and Germany. The data also shows that a good number of studies have been conducted in the South Korea (n = 8, 9.41) and Netherlands (n = 6, 7.06%). Our study infers that, compared to the 21 remaining countries, the above-mentioned countries invested a good amount of effort in domestic technological development towards developing XR-based mental disorder tools. Thus, XR systems contribute immensely to the economic and health systems of developed countries. Meanwhile, our data identified that such studies are not yet prioritized in Africa. In terms of study frequency by year, Table 2 shows that the majority of the articles 31.76% (n = 31) were published in the year 2021, i.e., of the total studies included. This could underline a global priority set to advance mental health care. However, none of the studies mentioned if they were motivated or attributed to the ongoing COVID-19 pandemic[122]. Nonetheless, secondary studies identified that prevalence of anxiety and depression increased by 25% in the first year of the pandemic, while the psychosocial effects of the pandemic varies by regions [123, 124]. Hence, it is likely that the pandemic-related increase in mental healths and the increased adoption on virtual treatment during the pandemic contributed to the rise in the number of XR-based mental health intervention in 2021. Furthermore, there was a great decline in the number of investigations reported in the years 2022 (14.12%) and 2023 (18.82%). This coincides with the time COVID-19's global prevalence had decreased.

## Demographics of XR Usage: Age Population

We analyzed the age of participants included in the studies as shown in Figure 4. We categorized the study participants as *Children*, *Teenagers*, *Adolescents*, *Young Adults*, *Young and Old Adults*, and *Old Adults* based on the age ranges reported in the 85 studies that were included in this review. We declared studies that omitted such information as "*Not Specified*". A substantial age overlap was found amongst the groups of individuals included in the reviewed studies. Thus, we classified ***Adults*** where participants ages were specified to be between 18 and 65 years old in the reported studies, and ***Older Adults*** where the participants are above 65 years old. As shown in Figure 4, we observed that half of the studies (n = 38, 44.71%) were designed for adults (people who are between 18 and 65 years old). We found that a small number of studies focused on younger age groups. For this age group, 10 of the 85 studies (11.76%) focused on participants between the age of 0 and 12 years old, while another two studies (2.35%) focused on groups of participants who are in their teenage years as well. The poor representation of participants from each age groups in the reviewed studies could be due to a lack of a standardized way for selecting a target audience when developing XR systems for managing mental disorders [125].

***Table 2.*** *Number of studies published per year*

| SN | Country of Study | Frequency | Percentage |
|---|---|---|---|
| 1 | 2016 | 3 | 3.53% |
| 2 | 2017 | 12 | 9.41% |
| 3 | 2018 | 5 | 4.71% |
| 4 | 2019 | 11 | 9.41% |
| 5 | 2020 | 10 | 8.24% |
| 6 | 2021 | 31 | 31.76% |
| 7 | 2022 | 12 | 14.12% |
| 8 | 2023 | 16 | 18.82% |

***Table 3.*** Audience group found in the included studies by level of maturity

| SN | Audience Group | Frequency | Age Range | Percentage | Participants | Percentage |
|---|---|---|---|---|---|---|
| 1 | Children | 10 | 1 – 10 Years | 27.06% | 1 – 10 | 11.76% |
| 2 | Teenagers | 2 | 11 – 20 Years | 55.29% | 11 – 20 | 10.59% |
| 3 | Adolescents | 27 | 21 – 60 Years | 8.24% | 21 – 50 | 25.88% |
| 4 | Adults | 38 | 61 – 85 Years | 1.18% | 51 – 100 | 27.06% |
| 5 | Older Adults | 3 | 85+ Years | 2.35% | >100 | 18.82% |
| 6 | Not Specified | 5 | Not Specified | 5.88% | Not Specified | 5.88% |

As analyzed in the third panel of Table 3, we indicate the common classification of participants' age rages that were reported in the selected studies. In addition to the statistical information derived from the age groups (as shown in the pie chart of Figure 4), This is because the age ranges used for defining the categories of participants' in the 85 articles were not unique. Also, there were great overlaps when comparing the category tags and age ranges across the different studies. Thus, we refined the data to synthesize mean age distribution of the participants included in each study. For this, the age ranges were set as given or generated as $\text{mean}_{\text{age}} - \text{SD} — \text{mean}_{\text{age}} + \text{SD}$ when only $\text{mean}_{\text{age}} \pm \text{SD}$ was given (SD implies standard deviation). The mean age distributions of the participants used for classification are reported in the second panel of Table 3. The data indicates that majority of the studies (n = 47, 55.29%) were designed for an audience with a mean age of $35.079 \pm 9.72$ years. The age distribution in this group is particularly dominated with lower and upper values of 18 years (n = 26/47) and 65 years (n = 7/47), respectively in the different participants' age ranges. The age range of youngest participants who were admitted to participated in the XR-based mental disorder study are a group of children between the age of 4 – 8 years old that investigated how VR reduces the perception of anxiety in infants [59].

### Demographics of XR Usage: Study Sample Size

In addition, we analyzed the sample size of participants admitted and found 85 studies meeting the inclusion criteria of this scoping review. We partitioned the sample size into five different categories and analyzed the number of studies as reported in the third panel of Table 3. It can be seen that the number of participants recruited for most studies were between 51 – 100 people (n = 23, 27.06%). Next to it are studies that recruited participants between 21 – 50 people (n = 22, 25.88%) to evaluate mental health with XR. Furthermore, 18.82% (n = 16) studies included over 100 participants, while a very small sample size (≤20 participants) were considered in 22.35% (n = 19) of the 85 studies. Most of the studies in the latter were more subtle in their findings and conclusions. Thus, it can be understood that having relatively more participants is helpful to reach better conclusions. Overall, each of the participant categories identified above were reported in at least 5 different studies. It is worth mentioning that only 5 studies (5.88%) did not specify the sample size of the participants used. The participants' gender distributions were not analyzed as this data was missing in most studies included in this scoping study.

### Demographics of Design and Implementation Strategies

Application of XR systems for mental disorders requires vigorous study and implementation strategies. We analysed the study different factors usually considered when designing or evaluating XR systems for mental health interventions. The three major considerations found in the selected studies are the *type* of the study performed, the *design* factors and the *evaluation method* used to assess each study. With a focus on anxiety and depression, the four main types of studies that were are carried out are i) "***discussion***" which are studies with a narrative focus; ii) "***experimental***": studies that were done to investigate the effect of the XR techniques on certain groups of subjects or other factors that aid or affect such setup; iii) "***modeling***" which includes studies that were done to develop new model or setup and such were validated on limited subjects or data; and lastly, iv) "***analysis***" which are the studies done without any particular experimental study but relying on data of previous ***experimental*** studies.

As reported in Table 4, it was found the 78.82% of the studies (n = 67) were frequently investigated experimentally to report how XR aids intervention of mental disorder. Meanwhile, studies based on ***modeling*** and ***analysis*** makes up of 9.41% (n = 8) of the included studies, each. Studies based on narration i.e. the ***Discussion*** category accounts for 1.18% of the 85 studies. This shows most studies were done under experimental investigation. Typically, this enables direct comparison between the mental conditions and relationships with their causal factors in psychological cornerstone studies [126]. Furthermore, 58.82% (n = 50) of the studies investigated the effects of XR immersion. This shows that researchers in this domain are commonly fond of investigating how immersion can influence the mental health procedures. The other design factors of the XR systems found in the 85 studies are on subject's process automation (n = 14, 16.47%). The latter were majorly investigated to observe if real-world situations and environment are well emulated in the studies. Similarly, cases of XR personalization and manual execution were typically focused in nine (10.59%) and six (7.06%) studies, respectively.

***Table 4.*** Demographics of the study implementation factors

| Type | Freq , Percent | Type | Freq , Percent | Evaluation Method | Freq, Percent |
|---|---|---|---|---|---|
| Discussion | 1, 1.18% | Personalization | 9, 10.59% | Quantitative Method | 32, 37.65% |
| Experimental | 67, 78.82% | Manual | 6, 7.06% | Qualitative Method | 26, 30.59% |
| Modeling | 8, 9.41% | Process Automation | 14, 16.47% | Mixed method | 23, 27.06% |
| Analysis | 8, 9.41% | Immersion Analysis | 50, 58.82% | Not Specified | 4, 4.71% |

## The Relationship between Study Periods and Duration per Session

In addition, we analyzed the common evaluation categories reported in the 85 studies. First, categories of study periods (in weeks) and the duration per session (in minutes) were analyzed with respect to the number of sessions in the studies. As presented in Figure 2a, most of the studies were carried out in 1 – 5 weeks (n = 25, 30.59%). For studies done in this time period, it was common to evaluate their participants under 1-5 sessions (n = 38, 44.71%). However, several other studies evaluated their participants within 6 – 10 sessions (n = 15, 17.65%) in studies that lasted 6 – 10 weeks (n = 11, 12.94%) in some cases. Also, a few studies were carried out under 11 – 15 weeks (n = 6, 7.06%). It is worth emphasizing that the longest study (n = 1) lasted for 16 – 20 weeks (1.18%).

Furthermore, we analyzed the duration per session (in minutes) for the sessions that were reported in each of the studies. In most of the studies (n = 45, 52.94%) participants used XR techniques for 0 – 30 minutes. This is followed by studies requiring 31 – 60 minutes (n = 19, 22.35%) and 61 – 90 minutes (n = 5, 5.88%) of user engagement per session. On the extreme end, XR technique was used for a single session that lasted over 150 minutes. It was observed that 13 studies (15.29%) did not specify the session durations.

In addition, we analyzed the common evaluation method used in most of the studies included in this review. Three main methods were identified, and these are quantitative, qualitative, and mixed evaluation methods. Overall, a qualitative assessment approach was applied in 51.67% (n = 31) of the 85 studies and it was understood that *qualitative method reveals deeper insights about the XR-based evaluation.* Furthermore, a quantitative method was used in 10 studies (16.67%), while a mixed methods approach was found in 17 studies (28.33%). We also found that the evaluation methods were not reported in 2 studies (see Figure 7c).

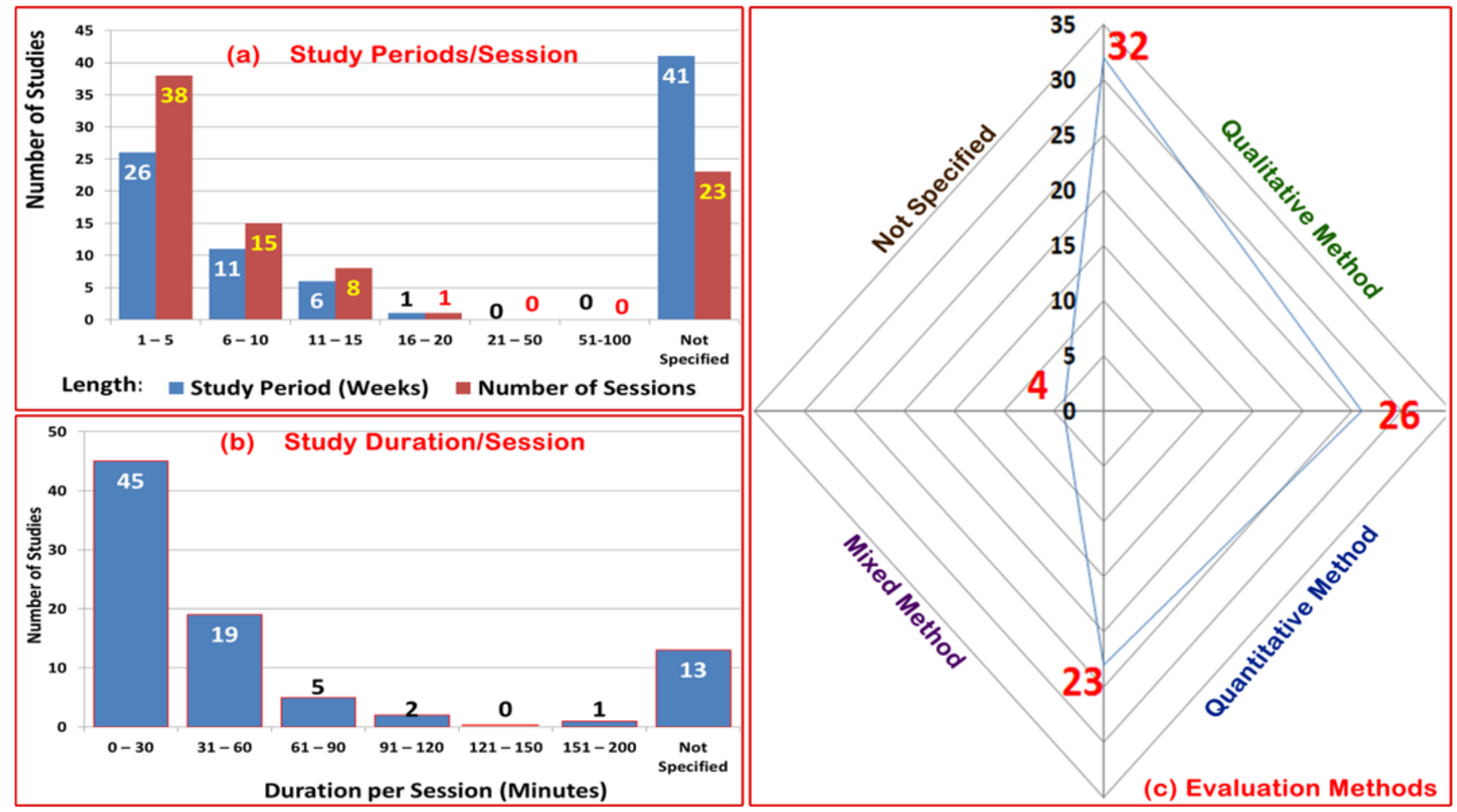

**Figure 2.** Categories of studies investigated for the different mental healths in the included studies

## Extended Reality and Gamification Strategy for Mental Disorders

It is important to analyse the XR techniques used for the intervention of depression and anxiety. First, we analyse the major strategies found in the studies. The XR tools in each study were identified to either be gamified or non-gamified. Gamification strategies were adopted in 30.59% (n = 26) of the studies, and these strategies were used across 10 mental disorders in the 85 articles. The few exceptions where a gamification strategy was not applied include *negative thoughts, panic disorder,* and *pain and anxiety*. Conversely, non-gamified strategies were adopted for the intervention of the remaining mental conditions. This accounts for 69.41% (n = 59) of the 85 studies. However, *alcohol use disorder* and *attachment behavior* were only addressed with gamified XR systems. It may be right to think that gamification strategies is yet to be matured for such condition, or possibly that the existing gamification strategies are not suitable when evaluating such mental disorders with XR techniques, or perhaps there are on-going studies to show its applicability.

## Extended Reality Development Tools

In addition, we analyzed the XR applications that have been used in the study and found that specific systems such as 3D Unity are commonly used by many authors in the development of their XR. Only 68 of the 85 studies (77.64%) reported the name of the actual XR app they implemented or adopted in their studies. As reported in Table 5, the most frequently tool for developing XR platforms is the 3D *Unity Pro* (n = 16, 23.53%). That is probably due to its powerful editor to create XR systems, and its support for cross-platform development. Similarly, we observed that several studies (n = 15, 22.06%) have been carried out with custom VR systems. Such adaptive applications are either newly developed or adopted in those studies and evaluated for aiding mental care. Meanwhile, *Blender* and Mobile virtual systems were used in 3 studies (4.41%) and 10 studies (14.71%), respectively. Another 24 studies (28.24%) indicated using a development platform but without specifying it, while the remaining 17 (20.0%) studies did not mention the use of development platform.

***Table 5.*** Software VR tools commonly used for XR development

| SN | Number of Participants | Studies (%) |
|---|---|---|
| 1 | 3D Unity Pro | 16 (18.82) |
| 2 | Custom adaptive VR software | 15 (17.64) |
| 3 | Blender 3D | 3 (3.52) |
| 4 | Mobile virtual system | 10 (11.76) |
| 5 | Others | 24 (28.24) |

*Table 6. Types of Hardware Technologies used for XR Intervention of Mental Disorders*

| SN | XR Technologies Used | Studies | SN | XR Technology Used | Number of Studies |
|---|---|---|---|---|---|
| 1 | VR Head-mount | 54% | 7 | Biopac MP150 | 3 % |
| 2 | 3D VR Glasses | 8% | 8 | Earbuds | 3 % |
| 3 | Smartphone | 7% | 9 | Location Trackers | 3 % |
| 4 | Google VR Box | 5 % | 10 | Directional Microphone | 1 % |
| 5 | EEG/EMG | 5% | 11 | Gamepad | 1 % |
| 6 | Headphone | 5% | 12 | Webcam | 1 % |

*Table 7. Types of VR Headsets*

| SN | Head Mounts Used | Number of Studies | SN | Head Mounts Used | Number of Studies |
|---|---|---|---|---|---|
| 1 | HTC Vive | 12 | 6 | Oculus Rift | 2 |
| 2 | Samsung Gear VR | 11 | 7 | Oculus CV1 | 1 |
| 3 | 3D VR Glasses | 7 | 8 | EEG VR Head | 1 |
| 4 | Oculus Go | 3 | 9 | Windows MR Headset | 1 |
| 5 | Google VR Box | 3 | 10 | | |

## Hardware Technologies used for XR in Mental Disorders

To further fulfil the aim of this study, we extracted information describing the XR technologies used to deliver the mental health care in the 85 studies. For the purposes of being as inclusive as possible, we only reported the hardware components that were listed for setting up the XR environments in the studies. Figure 10a shows that the use of headsets was consistent in 49.41% (n = 42) of the 85 studies thus, it is the most commonly used component when setting up XR for mental disorder intervention. Typically, it was found that Oculus *head mount display* and VR headsets were common in such studies. Also, use of smartphones is a common technology used in setting up the XR environment. It was found that 14.12% (n = 12) of the studies included smartphones of different types ion their studies. These hardware components (i.e. head mounts, smartphones and VR glasses) are increasingly popular in studies where gamification strategy was adopted. We further analyzed the most popular types of headsets and found that they are headphones, earbuds, and VR head mounts. The latter is a more advanced technology and a basic component in most XR studies. As reported in Table 7, there are nine different types of head mounts used in the 85 studies. HTC Vive and Samsung Gear VR was the most utilized head mount in setting up XR systems. These were found in 12 (14.12%) and 11 studies (12.94%) of the 85 studies. The next such head mount is the 3D VR glasses which were used in 7 studies (8.24%). In addition, the Oculus Go and Google VR Box were used in 3 studies (3.53%) each. Two articles reported to have used the Oculus Rift in their studies, while different types of VR simulators such as the Oculus CV1, a custom EEG cap with VR head display, and Windows MR headset (n = 1, 1.18%) each were also used.

## Extended Reality for Anxiety and Depressive Disorders

We identified 14 types of mental healths in the 85 articles and illustrated the count of each condition shown in Figure 9. We observed that most of the XR studies are centered around anxiety and depression (n = 53, 62.35%). These include the use of XR for anxiety without depression in 52.94% (n = 43) of the studies. These include studies where the primary and secondary focus includes anxiety, and depression was not considered at all. We found that minimal attention (n = 4, 4.71%) was on depression wherein anxiety of any kind was not considered. Typically, 27.05% (n = 23) of the studies focused on just anxiety, while the remaining 30 studies (35.29%) combined anxiety with other mental disorders. Anxiety and depressive disorders sometimes have very ambiguous borderline definitions; thus, this scoping review is more focused on them. We further expatiate by looking into individual conditions like social anxiety disorder and generalized anxiety disorder which were found in 16.47% and 3.53% of the studies (n = 14, n = 3), respectively. This condition was combined with unclassified distress in some studies [67, 127]. From the overall articles included in this review, 13 studies (10.0%) were found to have applied XR technologies for phobic related mental health (fear disorders that is a clinical evaluation of anxiety).

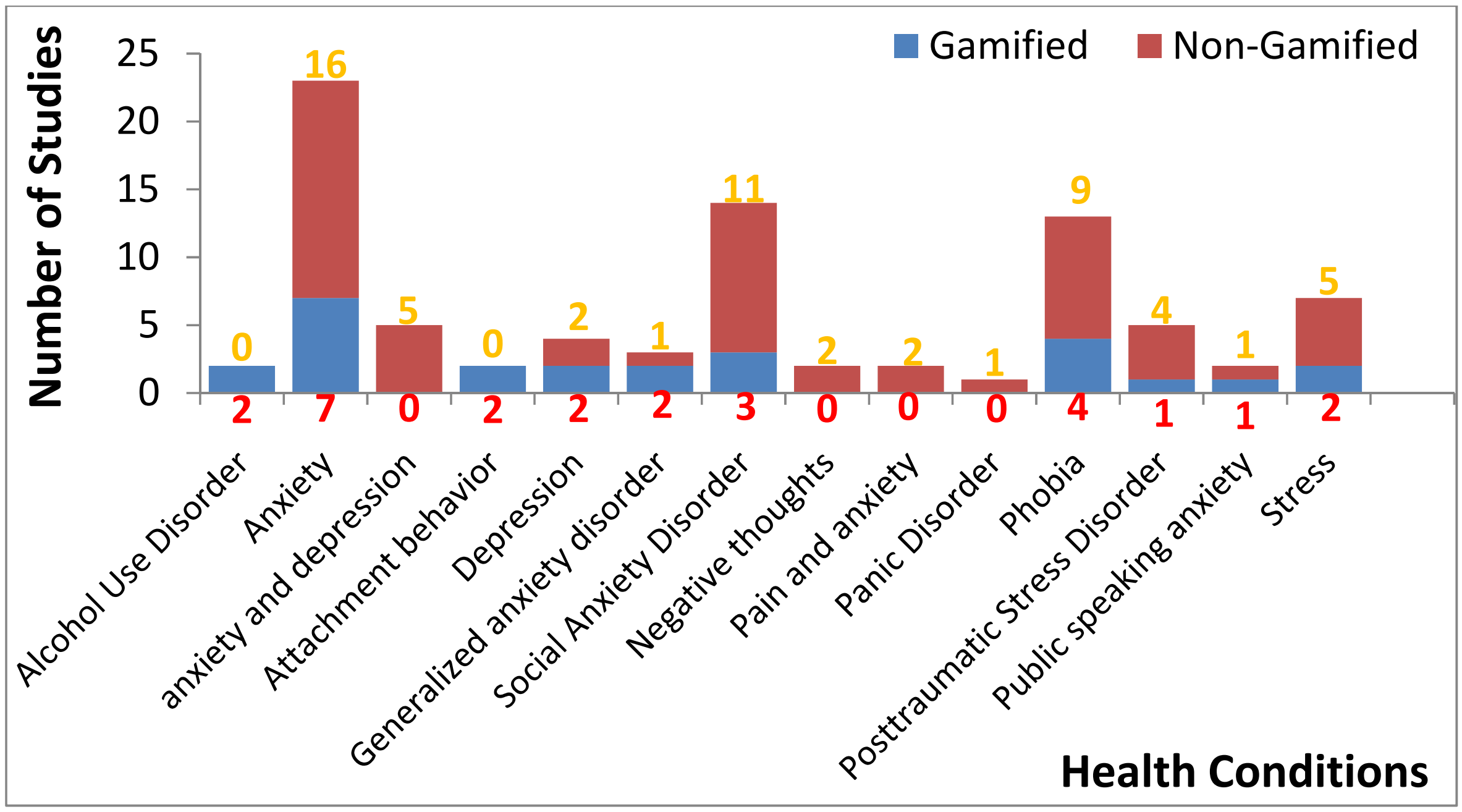

***Figure 3.*** *Number of studies per clinical condition*

Adoption of XR for other mental disorders without anxiety or depression was also studied. For instance, physiological disorders such as emotion and stress issues were investigated. Amongst these, phobia of different kinds e.g. Acrophobia, Claustrophobia, Fear were investigate in 15.29% (n = 13) of the studies, where posttraumatic stress disorder was studied in five articles. The latter has a similar frequency as one of the top mental disorder (i.e. depression). We also found that negative thoughts and attachment behavior were investigated in 2.35% of the studies (n = 2) each. In contrast to Baghaei *et al.'s* [128] findings, we found that generalized anxiety disorder was investigated as specific clinical conditions in 3.53% (n = 3) studies. Beyond this, the other mental disorders were either found in only two studies (i.e., 2.35%) such as in the case of alcoholic use disorder, attention disorder, attachment behavior, negative thoughts, Pain and anxiety, and public speaking anxiety; or only in one study (i.e., 1.67%) as the case of panic disorder. Thus, XR technology is commonly used for evaluating anxiety disorders. Finally, this scoping study shows that XR-based evaluations are distinctly applied for anxiety and other mental disorders that exclude depression. A typical case includes the development of XR system for anxiety and phobia, as well as anxiety and psychiatric disorders [88, 129]. The benefits of XR for evaluation and management of mental disorder was identified in the articles that were reviewed. Recent studies show that VR yields same level of effectiveness as exposure-based therapy for reducing anxiety symptoms [130]. This section mostly uncovers the use of VR technology in anxiety; however, it shows that AR and MR have been recently emphasized as an add-on technology and not a substitute. It becomes clear that more studies are still needed for evaluating how AR and MR can singly improve mental health.

## Discussion: Meta synthesis of the selected literature

In the last section, we focused on the demographics, technology, and study designs found in the existing XR systems used for mental disorder intervention. In this section, the effectiveness of the XR systems for anxiety and depression as top mental disorders are analysed as reported in the studies.

### Effectiveness of XR technology for mental disorder intervention

Following the review of the literature included in this scoping study, it can be concluded that XR systems are commonly used for managing mental disorders. In this scoping review, we found that the technologies have been majorly used for evaluating anxiety and depression separately, in combination with each other or other common mental disorders. For the latter, the majority of the studies were targeted at cognitive and behavioral change, i.e. subjective care to improve patients' behavior and/or attitude. Also, it was observed that of the XR technologies, VR-based systems are mostly used. For instance, it was showed in some studies [27, 78, 83] that VR could be effectively used to evaluate anxiety and depressive symptoms in patients with mental disorders. Similarly, Li & Luo [62] established that gamified XR can reduce depressive disorders through cognitive empathy and mutual understanding amongst patients and caregivers. Many studies reported that the XR systems help reduced the symptoms of the mental disorder focused. For instance, some authors [27, 42, 45] strongly indicated that using XR technology in psychotherapy process for reduced anxiety and depression in their subjects. Similarly, Niharika et al. [64] shows significant decrease in subjects' anxiety scores when using VR eyeglasses during dental treatment.

XR intervention is a safe noninvasive technique that does not require any previous education and training and has lasting effects. Referring to authors like McLay et al. [60], it was, however, seen that statistical significant differences between XR-based treatment method and the conventional approaches may not be a constant thing when applying XR systems for mental disorder interventions. In comparison to standard CBT, some authors [61, 74] improved the psychotherapy of depressive disorder in young adults by developing effective VR-enhanced personal construct therapy. Benjamin et al. [102], in the SoREAL study, investigated *in-vivo* group CBT and compared its effects with VR exposure CBT for patients diagnosed with social anxiety disorder.mental disorder Similarly, some authors Shin et al. [108] & Donker et al. [44] investigated the efficacy of mobile-based VR-CBTs for panic and phobic interventions. The app-based XR interventions were effective for managing the disorder symptoms, and restoring subjects' autonomic nervous system. This demonstrates the validity of using XR systems as self-guided and cost-effective therapeutic approach. Taken together, these studies show that recent development of XR technologies is gaining traction for mental disorder evaluation and treatment. Thus, some researchers suggested that future XR interventions should consider providing multi-user experiences that can help increase social engagements for patients that are possibly confined due to disabilities. Thus, it can be concluded that virtual environments are as effective as exposure therapy for evaluating mental health. We found studies investigating if gamified XR is also effective for reducing acrophobia. The stimuli presented using AR indeed induce physiological alterations in the participants [44, 131].

### Effectiveness of XR design factors on the outcomes of mental disorder interventions

Bras *et* al. [44] showed that AR and VR offer high levels of immersion and are optimal solutions for counteracting the effects of in-vivo exposure. Weerdmeester *et al.* [75] showed that, by engagement and cognitive biofeedback, gamified VR can reduce anxiety symptoms. Florian Grieger et al., investigate if XR-based interventions with multiple design factors can yield better results when used for intervention of mental disorders. In a random control trial study, authors found that personalized VRs aid a general positive shift in thoughts and emotions with increased relaxation and self-refection. This shows that VR systems with multiple features such as personalization, immersion and focus, interaction design and embodiment, integration can enhance treatment outcome. Similarly, Karlo Miguel et al., developed a mobile VR-based system for promoting relaxation to reduce anxiety and alter stressful activities among class students. Studies conducted by Fuad et al. [92] & Traister [121] show that XR technology can significantly lower students' anxiety, and as well enhanced them psychologically and physiologically with a safe and risk-free therapeutic experience. The related studies [57, 58] focused their works mental stress management in teenagers and adolescents. Brivio et al. [40], compared the efficiencies of 360° panorama and computer-simulated prototype in generating XR sense of presence, emotions, and relaxation when treating mental disorders. Also, Johan Lundin et al. [99], investigated whether filming virtual environments with a low-cost 360-degree film camera to produce VR-CBT can offer a feasible and acceptable treatment for some kinds of phobia. These studies show that virtual reality exposure therapy can produce long-lasting benefits for the mental disorders consistent with research on a variety of forms of short-term cognitive behavioral therapy for social anxiety disorder. The results showed that treatment satisfaction was high and participants had significant improvement at 6-month follow-up with large effect sizes. In another study, Veling *et al*. [88] showed that VR relaxation induces positive affective states and has short-term effects toward reducing psychiatric stress and anxiety symptoms disorder compared to standard relaxation exercises.

### Limitations

XR-based systems are posed to have some unique advantages over traditional methods used for mental disorder management. Nevertheless, XR systems also have some limitations. Future developments should consider technological innovation and standardization of treatment options. The following limitations should be considered when interpreting the results of this review. The developed search strategy was limited to using PubMed, Google Scholar and the ACM Digital Library databases for efficient and accurate search results. This may have excluded qualified articles from other databases. We also found that the number of weeks of evaluation for 41 studies (48.24%) and the number of sessions for 23 studies (27.06%) were not specified in the selected articles. Thus, it is hard to assert the best number of weeks and number of sessions needed for validating the use of XR-based technology in mental disorder evaluation. This study identified different major methodological approaches and development tool used by studies. Another limitation of the study is the lack of scientific assessment of the quality of the publications that were included in the scoping review. Moreover, due to the large number of articles reviewed, there is a possibility of overlooking valid publications that might have met the inclusion criteria. Non-English articles were not included in this review either. Finally, considering the possibility of bias in the reported outcomes for many reasons including due to self-reporting and publishing bias that tend to favour papers with positive outcomes, the findings of this scoping review should be applied with caution especially regarding the effectiveness of XR-based intervention for mental disorder.

## Conclusion and Future Work

Extended reality therapy has been widely used for a variety of mental health care. This scoping review investigated the adoption of XR in mental disorder, specifically, anxiety and depression. The review covered 85 studies which used different types of VR, AR, and MR technologies for mental disorder with a focus on anxiety and depression. The study uncovered that majority of reviewed articles reported a reduction in symptoms of anxiety or depression with the use of XR. Moreover, the studies adopted unique designs that are set up to monitor the signs of the mental health. The recorded signs can be used for formulating appropriate therapy. We also found that XR-based interventions have been shown to be effective approaches with a high level of users' acceptability in 18 mental health conditions. Although a considerable number of studies (n = 85) were included in this scoping review, some areas are still under-researched hence not well-represented in the reviewed studies. For instance, adoption of non-gamified strategies was found to have cut across 18 mental health conditions included in this review study. However, studies investigating pain and anxiety, negative thoughts, autoimmune disorder, and acquired brain injury did not employ any form of gamification strategies. This study was done to investigate the implementation and adoption levels of XR for mental health care delivery. Our study outputs indicate that many studies focused on anxiety either alone, or in combination with other conditions. Meanwhile, a limited number of studies found to solely focus on depression. In a previous study, Baghaei *et al.* [128] also showed that supporting people with depression in XR settings is an interesting area to explore for mental health care. As presented in *the Supplementary Materials*, only of the included studies carried out randomized controlled trials to study and compare the effectiveness of XR techniques. Thus, we recommend that future work should conduct controlled trials to investigate and compare the effectiveness of using XR-based intervention in mental health care and the benefits and costs of XR in the mental disorder management.

### Acknowledgements

Nilufar Baghaei acknowledges the University of Queensland for the Startup Grant. Olatunji Omisore was supported by the Shenzhen Natural Science Foundation under the Grant #JCYJ20190812173205538. Rita Orji acknowledges funding support from the Canada Research Chairs (CRC) Program and the Natural Sciences and Engineering Research Council of Canada (NSERC) through the Discovery Grant.

## Data sharing statements

The data used for this review study has been included as supplementary information files with the manuscript.

## Conflicts of Interest

None declared.

## Multimedia Appendices

Appendix 1: A summary of reviewed articles with extracted information

Appendix 2: The PRISMA checklist followed